\documentclass[%
 reprint,
longbibliography,
 amsmath,amssymb,
 aps,
 pra,
floatfix,
]{revtex4-2}

\usepackage[separate-uncertainty=true]{siunitx}

\usepackage{graphicx}% Include figure files
\usepackage{epstopdf}
\usepackage{gensymb}
\usepackage{dcolumn}% Align table columns on decimal point
\usepackage{bm}% bold math
\usepackage{color}
\usepackage{ulem}%allows strikethrough via \sout{}
\usepackage[dvipsnames]{xcolor} %this package provides more colours
\usepackage{caption}
\usepackage{subcaption}
\usepackage{hyperref}% add hypertext capabilities
\usepackage{soul}
\usepackage{booktabs}
\hypersetup{colorlinks=true,linkcolor=blue,filecolor=magenta,urlcolor=blue,citecolor=blue}
\begin{document}

\preprint{APS/123-QED}

\title{Characterisation and mitigation of RF knockout during beam stacking}% Force line breaks with \\

\author{
C.~Jolly, 
D.~J.~Kelliher, 
J.-B.~Lagrange, 
A.~P.~Letchford, 
S.~Machida,\\ 
D.~W.~Posthuma de Boer, 
C.~T.~Rogers, 
A.~Seville}

\affiliation{
 STFC ISIS Department, Rutherford Appleton Laboratory,\\
 Harwell Campus, Didcot, OX11 0QX United Kingdom% with \\
}%

\date{\today}% It is always \today, today,

\begin{abstract}

Beam stacking allows a Fixed Field alternating gradient Accelerator (FFA) to increase the extracted beam current whilst also allowing for a flexible time structure making FFAs a promising candidate for future spallation neutron sources and high beam intensity applications. For successful beam stacking, beam loss caused by RF knockout must be avoided. RF knockout can occur during beam stacking because of the finite dispersion function at the RF cavity location, which is unavoidable in a scaling FFA. In this work, the RF knockout resonance is characterised and through a series of experiments at the ISIS Neutron and Muon Source, we show that it is possible to suppress the loss from RF knockout.

\end{abstract}

\maketitle

\section{\label{sec:introduction} Introduction}

Beam stacking is one of the key advantages of Fixed Field alternating gradient Accelerators (FFAs) for high intensity applications and could be a crucial technique in an FFA-based spallation neutron source \cite{FFA_CDR_ePub}. With stacking, the beam can approach the space-charge limit at the extraction energy, allowing for a highly intense single pulse with greater flexibility in the time structure, when compared with conventional synchrotron designs.

\begin{figure}[htb]
    \centering
    \includegraphics[width=\columnwidth]{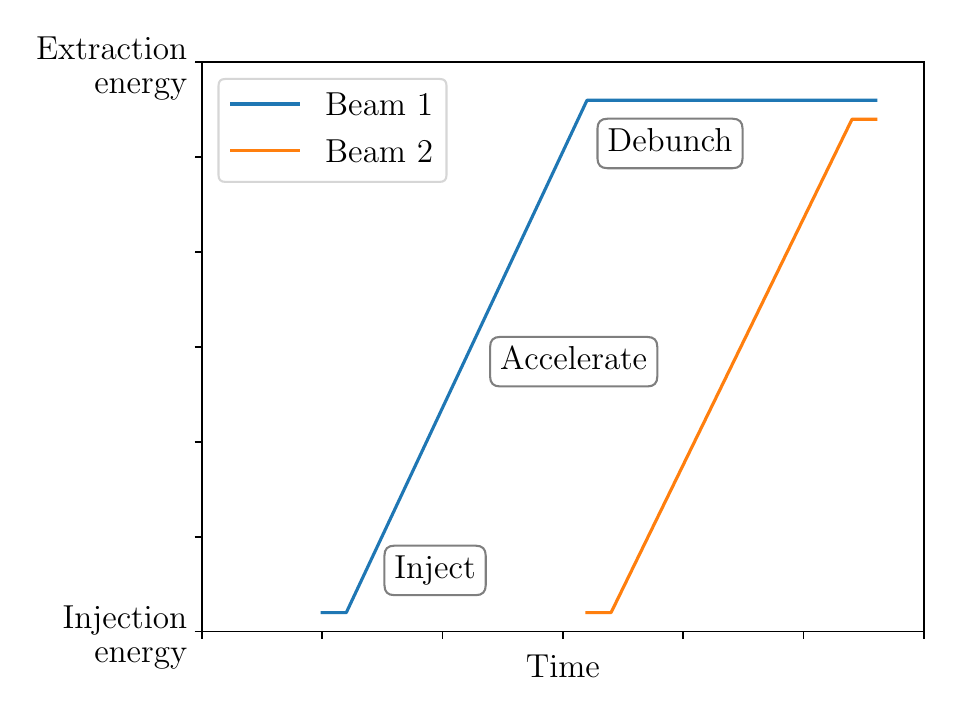}
    \caption{Shows an example beam stacking RF programme with linear RF ramps. At the end of this programme, two coasting beams will have been stacked at the extraction orbit.}
    \label{fig:beam_stacking_example_program}
\end{figure}

First achieved at MURA \cite{BeamStackingRSI}, beam stacking was later systematically demonstrated at Kyoto University's KURNS FFA \cite{BeamStackingPaper_PhysRevAccelBeams}. While two beams were successfully stacked at the KURNS FFA, significant beam loss was observed during the stacking process. An RF programme for beam stacking is shown in Fig.~\ref{fig:beam_stacking_example_program}. During stacking the first beam is injected and accelerated as usual. The beam is then debunched and stored as a coasting beam at the extraction energy. A second beam is then injected and accelerated. In this phase, the second beam (shown in orange in Fig.~\ref{fig:beam_stacking_example_program}) is accelerated whilst the first beam is stored. Finally, the second beam is brought to an energy just less than that of the first beam and debunched, leaving two coasting beams that can now be thought of as a single stacked beam.

In a scaling FFA, the dispersion function is finite all around the ring so that the orbits of different momenta become photographic enlargements. Crucially, when a particle in the stored beam passes a cavity and gains or loses energy, the position of the equilibrium orbit changes corresponding to the new momentum. Since the particle is now offset with respect to the new equilibrium orbit this causes a change of the betatron oscillation amplitude. Since the stored beam is not synchronised with the accelerating RF field of the second beam during stacking, the displacements nominally average out to zero.

However, when the RF frequency---which determines the frequency of the displacement---synchronises with the phase advance of a betatron oscillation between cavities,
the displacements could build up turn-by-turn,

in a similar fashion to a transverse integer resonance.

This effect is known as RF knockout and is consistent with the cause of beam loss during the beam stacking demonstration experiment at the KURNS FFA in 2023~ \cite{BeamStackingPaper_PhysRevAccelBeams}. The loss from RF knockout must be avoided for beam stacking to be a viable technique at a high intensity FFA. 
In this work, RF knockout is characterised and mitigation methods are tested experimentally at the ISIS Neutron and Muon Source \cite{Thomason:2019fwe}.

\section{\label{sec:RF knock-out} RF knockout}

\subsection{RF knockout mechanism}

RF knockout is an external perturbation to a beam circulating in an accelerator. For example, a transverse RF field with a certain frequency can excite a coherent transverse beam oscillation when driven at the betatron frequency. An RF cavity can be a source of RF knockout when it is placed at a location with a finite dispersion function.

In a scaling FFA \cite{Symon_FFA,okawaffa,ffa_e}, 
the equilibrium orbit radius is determined by the momentum,
\begin{align}
    \frac{r_0+\delta x}{r_0}=\left( \frac{p_0+\delta p}{p_0} \right)^\frac{1}{k+1},
    \label{eqn:FFA_orbit_radius}
\end{align}
where $r_0$ is the machine radius, $p_0$ is the beam momentum (the zero subscript indicates a reference value), $k$ is the geometric field index, $\delta x$ and $\delta p$ are the change in horizontal position and beam momentum respectively. Taking the derivative of Eq.~(\ref{eqn:FFA_orbit_radius}) gives the dispersion function, $D_x$, in a scaling FFA;
\begin{align*}
D_x &= \frac{r_0}{k+1}.
\end{align*}
An increase of the particle energy at the RF cavities suddenly moves the equilibrium orbit.
The magnitude of the displacement is proportional to the energy change, $\delta E$, and the dispersion function at the cavity,
\begin{align*}
\delta x &= -\frac{\gamma D_x}{1+\gamma}\frac{\delta E}{E}.
\end{align*}
where $\gamma$ is the Lorentz factor and $E$ is the kinetic energy.
The minus sign indicates that the closed orbit position moves, not the orbit.

The cumulative effect of the displacements over many turns can be written as a time-dependent perturbation modulated by the cavity frequency $\omega_\textrm{RF}$.

\begin{align}
    \delta x\left( t \right) 
    &=-\frac{\gamma D_x}{1+\gamma}
    \frac{V_\textrm{RF}\sin\left( \omega_\textrm{RF}t+\phi \right)}{E}
    \sum_{j=-\infty}^{\infty} \delta\left(\frac{t}{T_\textrm{rev}}-j\right) \notag \\
    &=-\frac{\gamma D_x}{ 1+\gamma}
    \frac{V_\textrm{RF}\sin\left( \omega_\textrm{RF}t+\phi \right)}{E}
    \sum_{n=-\infty}^{\infty}
    e^{ i n\omega_\textrm{rev} t} \notag \\
    &=\frac{i\gamma D_xV_\textrm{RF}}{2\left( 1+\gamma \right)E
    } \notag \\
    &\times \sum_{n=-\infty}^{\infty}
    \left( 
    e^{ i \left( \omega_\textrm{RF}+n\omega_\textrm{rev} \right) t+i\phi}
    - e^{ i \left( -\omega_\textrm{RF}+n\omega_\textrm{rev} \right) t-i\phi}
    \right)
\label{eq:deltax}
\end{align}
where $j$ is the index of the RF knockout kick, $t$ time,  $T_\textrm{rev}$ the revolution time, $V_\textrm{RF}$ the RF voltage, $\phi$ the cavity phase offset and $i$ the imaginary unit. $n$ will be defined as the mode number at the end of this subsection.

The displacement to a particle when passing the cavity is represented by the sum of delta functions; it is convenient to replace the sum of delta functions with a sum of complex exponentials using a Fourier series.

A Hamiltonian for the beam equation of motion including the displacements from RF knockout can be constructed as,

\begin{align*}
    H=\frac{p_x^2}{2m}+\frac{1}{2}m\omega_{\beta}^2x^2-\frac{\delta x}{T_\textrm{rev}}p_x,
\end{align*}
where $m$ the particle mass, $p_x$ is the horizontal canonical momentum, $\omega_\beta$ is the betatron oscillation angular frequency.
The equations of motion are;
\begin{align*}
    \frac{dx}{dt}&=\frac{\partial H}{\partial p_x}=\frac{p_x}{m}-\frac{\delta x}{T_\textrm{rev}}, \\
    \frac{dp_x}{dt}&=-\frac{\partial H}{\partial x}=-m\omega_\beta^2 x,
\end{align*}
and
\iffalse
\begin{align*}
    \frac{d^2 x}{dt^2}=\frac{1}{m}\frac{dp_x}{dt}-\frac{1}{T_\textrm{rev}}\frac{d\delta x}{dt}.
\end{align*}
or
\fi
\begin{align*}
       \frac{d^2 x}{dt^2}+\omega_\beta^2 x=-\frac{1}{T_\textrm{rev}}\frac{d\delta x}{dt}.
\end{align*}
\iffalse
\begin{align*}
    p_x&=\frac{dx}{dt}+\frac{1}{T_\textrm{rev}}\int^t\delta x(t_1)dt_1, \\
    \frac{dp_x}{dt}&=-\omega_\beta^2 x.
\end{align*}
\fi

Therefore,
\begin{align}
    \frac{d^2 x}{dt^2}+\omega_\beta^2x = \frac{\gamma D_xV_\textrm{RF}}{2\left( 1+\gamma \right)ET_\textrm{rev}} \notag \\
    \times \sum_{n=-\infty}^{\infty}
    \left( 
    C_{+}e^{ i \left( \omega_\textrm{RF}+n\omega_\textrm{rev} \right) t}
    - C_{-}e^{ i \left( -\omega_\textrm{RF}+n\omega_\textrm{rev} \right) t}
    \right),
    \label{eqn:single_cavity_force}
\end{align}
\iffalse
\begin{align}
    \frac{d^2 x}{dt^2}+\omega_\beta^2x = \frac{iD_xV_\textrm{RF}}{2\left( 1+\gamma \right)ET_\textrm{rev}} \notag \\
    \times \sum_{n=-\infty}^{\infty}
    \left( 
    e^{ i \left( \omega_\textrm{RF}+n\omega_\textrm{rev} \right) t +i\phi}
    - e^{ i \left( -\omega_\textrm{RF}+n\omega_\textrm{rev} \right) t -i\phi}
    \right),
    \label{eqn:single_cavity_force}
\end{align}
\fi
and
\begin{align*}
    C_{+}&=\left( \omega_\textrm{RF}+n\omega_\textrm{rev} \right)e^{i\phi}, \\
    C_{-}&=\left( -\omega_\textrm{RF}+n\omega_\textrm{rev} \right)e^{-i\phi}. \\
\end{align*}

This will drive a betatron resonance when,
\begin{equation}
    \omega_\beta = \pm \omega_\textrm{RF} + n\omega_\textrm{rev}.
        \label{eq:resonancecond}
\end{equation}
By dividing through by the revolution frequency $\omega_\textrm{rev}$,
the RF knockout condition can be written in terms of the horizontal tune $Q_x$, 
\begin{equation*}\label{eq:knockout1}
    Q_x = \pm \frac{\omega_\textrm{RF}}{\omega_\textrm{rev}}+ n.
\end{equation*}
The integer $n$ can be interpreted as the mode number of the standing wave around the circumference, produced by RF knockout.

\subsection{Amplitude growth from RF knockout}

Following the analysis in \cite{Terwilliger_mura133}, we derive the growth in the betatron amplitude due to RF knockout
when the RF frequency is swept.
We assume that the resonance condition of Eq.~(\ref{eq:resonancecond}) is satisfied at some point during ramping of the RF frequency.
Regardless of the actual pattern of an RF frequency programme, we can approximate that the RF frequency change is linear in time around the period when the resonance occurs;

\begin{align*}
    \omega_\textrm{RF}\left( t \right) &= \omega_\beta - n\omega_\textrm{rev} + \alpha t,
\end{align*}
where $\alpha$ is the gradient of the linear frequency ramp. With a small phase difference of $\Omega\left( t \right)$ between betatron oscillations and the displacement,
\begin{align*}
    \Omega \left( t \right)&=\int_0^t \left[ \left(\omega_\textrm{RF} \left( t \right)+n\omega_\textrm{rev}\right)-\omega_\beta \right]dt \\
    &=\int_0^t \alpha tdt = \frac{\alpha}{2}t^2,
\end{align*}
one frequency component of Eq.(\ref{eq:deltax}) is cumulatively added 
\begin{align*}
    \int_{-\infty}^\infty \delta x\left(t\right)dt
    &= \frac{if\gamma D_xV_\textrm{RF}}{2\left( 1+\gamma \right)E}
    \int_{-\infty}^\infty e^{i\Omega\left( t \right)}dt \\
        &= \frac{if\gamma D_xV_\textrm{RF}}{2\left( 1+\gamma \right)E}
    \sqrt{\frac{2\pi}{\alpha}}e^{i\pi/4}, \\
\end{align*}
where $f$ is the number of RF knockout displacements per second.
The absolute value gives the amplitude growth,
\begin{align}
    \left| \int_{-\infty}^\infty \delta x\left(t\right)dt \right|
    &= \frac{f\gamma D_xV_\textrm{RF}}{2\left( 1+\gamma \right)E}\sqrt{\frac{2\pi}{\alpha}} .
\label{eqn:RFKO_amplitude}
\end{align}

From Eq.~(\ref{eqn:RFKO_amplitude}), the amplitude increase from passing through an RF knockout resonance is expected to vary linearly with voltage and the inverse square root of the crossing rate. We can arrive at a similar result by analogy with an integer resonance crossing. R. Baartman \cite{RBaartman-resonance-crossing} and J. Le Duff \cite{LeDuff:1979zz} give equations with a similar form to Eq.~(\ref{eqn:RFKO_amplitude}), where the amplitude from an integer resonance crossing is linear with the magnetic field strength and the inverse square root of the crossing speed.

\subsection{Mitigation methods}

RF knockout mitigation methods were explored extensively at MURA \cite{kerst_suppression_1957, jones_experiments_1957}, here we review and test the mitigation methods experimentally. 

\begin{figure}[htb]
    \centering
    \includegraphics[width=\columnwidth]{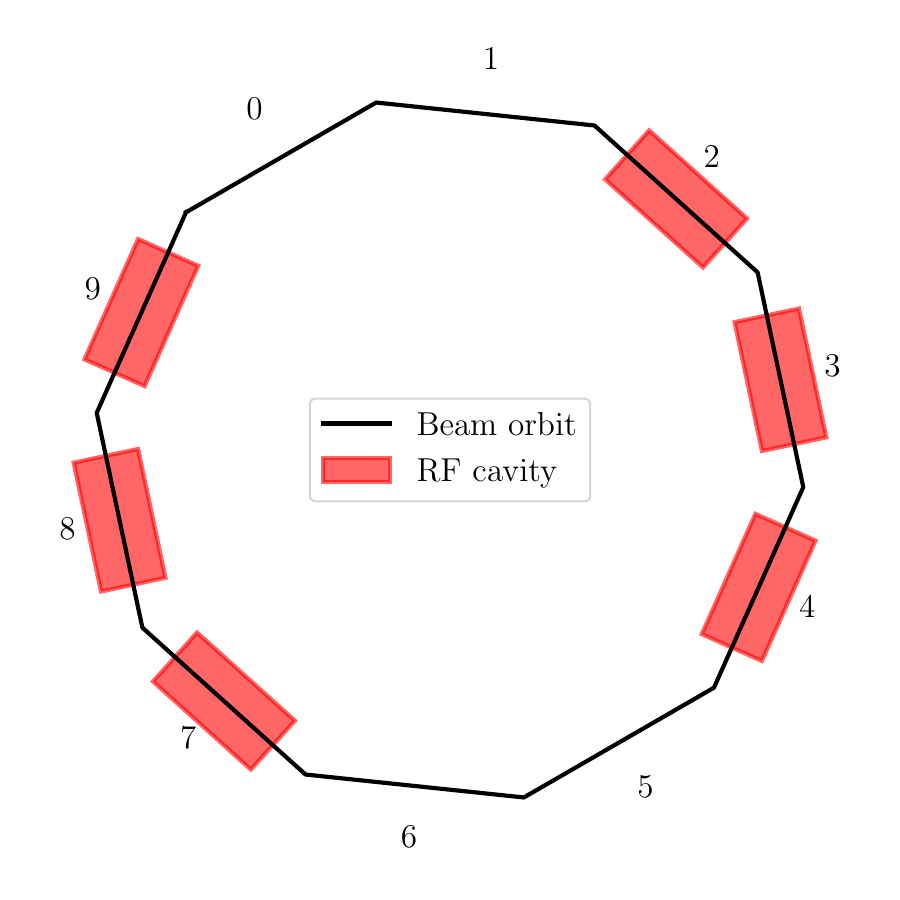}
    \caption{A diagram of the fundamental (harmonic 2) RF cavity positions in the ISIS ring, the superperiods are numbered from $0$ to $9$. Each cavity is located at the same point in the superperiod.}
    \label{fig:ISIS_h=2_cavity_diagram}
\end{figure}

The ISIS proton accelerator is conveniently arranged for testing RF knockout. ISIS's six fundamental (harmonic number $h=2$) cavities are positioned in three consecutive superperiods in two groups on opposite sides of the ring \cite{Barratt:963925}. The cavity arrangement is shown in Fig.~\ref{fig:ISIS_h=2_cavity_diagram}. Crucially, the cavities are located at the same position in each cell so the dispersion function is the same at each cavity. By powering only some of the six cavities, several arrangements can be tested. In this study, we consider two cancellation methods, a local cancellation and a global cancellation of the RF knockout displacements.

\subsubsection{Local cancellation}

A sudden change in the equilibrium orbit due to the energy gain in the RF cavity and the finite dispersion function at the cavity can be cancelled by other cavities located downstream.

\begin{figure}[htb]
    \centering
    \includegraphics[width=\columnwidth]{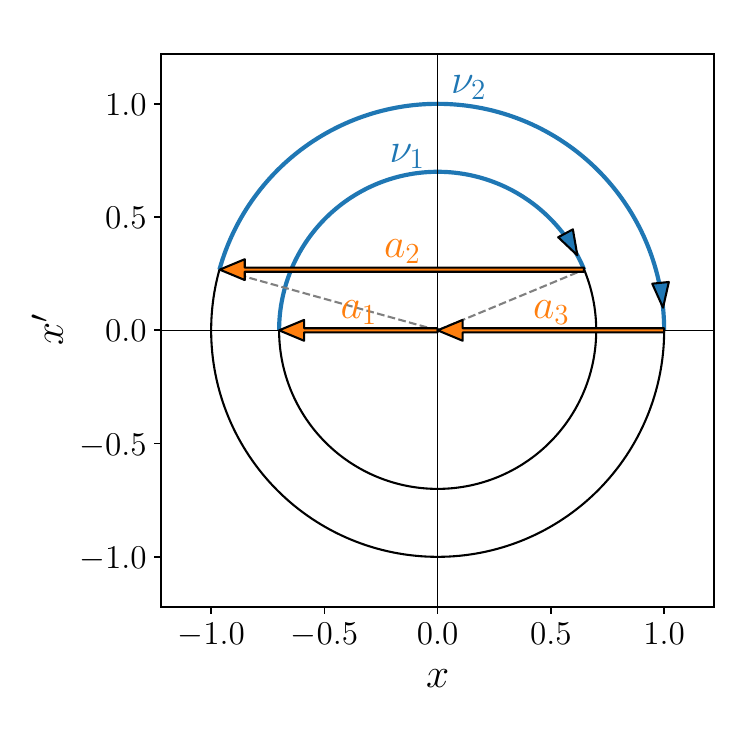}
    \caption{A diagram showing how the displacements from RF knockout are cancelled in a single turn using 3 successive cavities. $a_1, a_2$ and $a_3$ are the displacements at each cavity, $\nu_1$ and $\nu_2$ is the phase advance between the cavities.}
    \label{fig:3_cavity_local_cancellation}
\end{figure}

We assume that the amplitude of the dispersion function at two cavities is the same and energy gains are the same as well. 
Figure~\ref{fig:3_cavity_local_cancellation} shows a diagram of how the RF knockout displacements could be cancelled in a single turn by adjusting the relative amplitudes ($a_1, a_2$ and $a_3$) of 3 successive cavities. The 3 cavities are each separated by one superperiod. The phase advance from one cavity to the next is represented by $\nu_1$ and $\nu_2$. Starting from the origin, a particle experiences a displacement $a_1$ from the first cavity, then moves through a betatron phase advance, $\nu_1$, to the next cavity which has been set to a greater voltage. The particle experiences a correspondingly greater displacement from RF knockout. Finally, the particle moves though another phase advance, $\nu_2$, to the final cavity where it is given a final displacement $a_3$ to bring the particle back to the origin, thereby cancelling the RF knockout displacements in a single turn. 

An experimental test of mitigating RF knockout by locally cancelling the displacements is discussed in Sec.~\ref{sec:local_cancellation}.

\subsubsection{\label{sec:mitigation_with_symmetric_cavities}Global cancellation with symmetric cavities}

Following~\cite{Terwilliger_mura133}, we will derive the conditions of the global correction. By adding an extra term, the driving force in Eq.~(\ref{eqn:single_cavity_force}) can be modified to account for $M$ cavities spread throughout the ring. Writing the location of the $m^{\text{th}}$ cavity as $\theta_m$, the azimuthal position with respect to a nominal fixed point is;
\begin{equation*}
    \theta_m = 2 \pi \frac{L_m}{C},
\end{equation*}
where $L_m$ is the cavity location in the ring and $C$ is the ring circumference.

Eq.~(\ref{eq:deltax}) for $M$ cavities is modified as

\begin{align*}
\delta x\left(t\right) =A \sum _{m=1}^M 
\sin\left(\omega_\textrm{RF}t-h\theta_m+\phi\right) \\
\times \sum_{j=-\infty}^\infty \delta (\frac{t}{T_\textrm{rev}} - \frac{\theta_m}{2\pi} - j),
\end{align*}
where $A$ is the time independent constant of Eq.~(\ref{eq:deltax}),
\begin{align*}
    A=-\frac{\gamma D_x}{1+\gamma}
    \frac{V_\textrm{RF}}{E},
\end{align*}
$h$ is the harmonic number of the RF system. The $h\theta_m$ term accounts for the RF phase delay between different cavity positions while the $\frac{\theta_m}{2\pi}$ term accounts for the extra kick the particle receives from the $m^{\text{th}}$ cavity.

Following the same procedure as the single cavity case, the driving force can be written as a sum of exponential functions:

\begin{align*}
    \delta x(t) 
    &=\frac{iA}{2}  
    \sum_{n=-\infty}^\infty e^{i(\omega_\textrm{RF}+ n  \omega_\textrm{rev})t+i\phi}\sum_{m=1}^M e^{-i \theta_m(h + n)}, \\
    &-\frac{iA}{2}  
    \sum_{n=-\infty}^\infty e^{i(-\omega_\textrm{RF}+ n  \omega_\textrm{rev})t-i\phi}\sum_{m=1}^M e^{-i \theta_m(-h + n)},
\end{align*}

with the time dependent and independent terms separated into two exponentials.

The driving force from RF knockout will be zero when;
\begin{align}
\label{eqn:phasesum1}
\sum_{m=1}^M e^{-i\theta_m\left( h+n \right)}&=0
\textrm{    for }
\omega_\beta=\omega_\textrm{RF}+n\omega_\textrm{rev} \\
\label{eqn:phasesum2}
\sum_{m=1}^M e^{-i\theta_m\left( -h+n \right)}&=0
\textrm{    for }
\omega_\beta=-\omega_\textrm{RF}+n\omega_\textrm{rev}
\end{align}

Consider two special cases: one with equally positioned RF cavities around the ring, and one with two cavities positioned with an azimuthal angle difference of $\Delta\theta$.

When multiple cavities are equally positioned around the ring, $\theta_m$ becomes;
\begin{align*}
\theta_m \rightarrow 2 \pi \frac{m}{M},
\end{align*}

For this special case, Eq.~(\ref{eqn:phasesum1}) can be evaluated using the standard result for a geometric series;
\begin{align*}
\sum_{m=1}^{M} e^{-i2\pi \frac{m \left( h+n \right)}{M}}
=\frac{1-e^{-i2\pi \left( h+n \right)}}{1-e^{-i2\pi \frac{\left( h+n \right)}{M}}}.
\end{align*}

The sum is non-zero if;
\begin{align*}
\frac{h+n}{M}=p,
\end{align*}
where $p$ is a positive or negative integer.
When this condition is satisfied,
\begin{align*}
\sum_{m=1}^{M} e^{-i2\pi \frac{m\left( h+n \right)}{M}}
=M.
\end{align*}
Therefore,
the condition to avoid the RF knockout is,
\begin{align}\label{eq:mitigation11}
\frac{h+n}{M} \neq p
\textrm{    for }
\omega_\beta=\omega_\textrm{RF}+n\omega_\textrm{rev}.
\end{align}
Similarly, from Eq.~(\ref{eqn:phasesum2}),
\begin{align}\label{eq:mitigation12}
\frac{-h+n}{M} \neq p
\textrm{    for }
\omega_\beta=-\omega_\textrm{RF}+n\omega_\textrm{rev}.
\end{align}
It is clear that the number of knockout conditions is reduced proportionally to the number of RF cavities $M$.
Table~\ref{tab:condition1} shows how the symmetrically positioned RF cavities suppresses some of the RF knockout conditions of Eqs.~(\ref{eq:mitigation11}) and (\ref{eq:mitigation12}).
The parameters are based on the ISIS accelerator where $Q_x={\omega_\beta}/{\omega_\textrm{rev}}$=4.316 and $h=2$.

\begin{table}[htb]
\caption{$\pm \omega_\textrm{RF}/\omega_\textrm{rev}$ vs $M$. When $\frac{h+n}{M}$ or $\frac{-h+n}{M}$ is not an integer, RF knockout is mitigated. Otherwise RF knockout occurs indicated with a red value.}
\label{tab:condition1}
\begin{ruledtabular}
\begin{tabular}{cccccc}
$\omega_\textrm{RF}/\omega_\textrm{rev}$ & n 
& $M=1$
& $M=2$
& $M=3$
& $M=4$
\\
\colrule
0.316 & 4 & \textcolor{red}{6} & \textcolor{red}{3} & \textcolor{red}{2} & 6/4\\
-0.684 & 5 & \textcolor{red}{3} & 3/2 & \textcolor{red}{1} & 3/4\\
1.316 & 3 & \textcolor{red}{5} & 5/2 & 5/3 & 5/4\\
-1.684 & 6 & \textcolor{red}{4} & \textcolor{red}{2} & 4/3 & \textcolor{red}{1}\\
2.316 & 2 & \textcolor{red}{4} & \textcolor{red}{2} & 4/3 & \textcolor{red}{1}\\
-2.684 & 7 & \textcolor{red}{5} & 5/2 & 5/3 & 5/4\\
3.316 & 1 & \textcolor{red}{3} & 3/2 & \textcolor{red}{1} & 3/4\\
-3.684 & 8 & \textcolor{red}{6} & \textcolor{red}{3} & \textcolor{red}{2} & 6/4\\
4.316 & 0 & \textcolor{red}{2} & \textcolor{red}{1} & 2/3 & 2/4\\
-4.684 & 9 & \textcolor{red}{7} & 7/2 & 7/3 & 7/4\\
5.316 & -1 & \textcolor{red}{1} & 1/2 & 1/3 & 1/4\\
-5.684 & 10 & \textcolor{red}{8} & \textcolor{red}{4} & 8/3 & \textcolor{red}{2}\\
6.316 & -2 & \textcolor{red}{0} & \textcolor{red}{0} & \textcolor{red}{0} & \textcolor{red}{0}\\
-6.684 & 11 & \textcolor{red}{9} & 9/2 & \textcolor{red}{3} & 9/4\\
\end{tabular}
\end{ruledtabular}
\end{table}

When there are two cavities with the azimuthal angle difference of $\Delta\theta$, Eq.~(\ref{eqn:phasesum1}) becomes,
\begin{align}
\sum_m e^{-i \theta_m\left( h+n \right)}=1+e^{-i \Delta\theta\left( h+n \right)},
\label{eqn:opposite_cavities_time_indep_term}
\end{align}
where we assume that the first cavity is positioned at a nominal fixed point $\theta_{m=1}=0$ and the second cavity with an azimuthal position $\theta_{m=1}=\Delta\theta$.
The condition to avoid the RF knockout is then,

\begin{align}
\label{eq:mitigation21}
\Delta\theta\left( h+n \right)=\pm \left( 2d - 1 \right)\pi
\textrm{    for }
\omega_\beta=\omega_\textrm{RF}+n\omega_\textrm{rev}, \\
\label{eq:mitigation22}
\Delta\theta\left( -h+n \right)=\pm \left( 2d - 1 \right)\pi
\textrm{    for }
\omega_\beta=-\omega_\textrm{RF}+n\omega_\textrm{rev},
\end{align}
where $d$ is an integer.

\begin{table}[htb]
\caption{$\pm \omega_\textrm{RF}/\omega_\textrm{rev}$ vs $\Delta\theta/2\pi$. When $\Delta\theta (h+n)/\pi$ or $\Delta\theta (-h+n)/\pi$ is an odd integer number, RF knockout is mitigated. Otherwise RF knockout occurs indicated with a red value.}
\label{tab:condition3a}
\begin{ruledtabular}
\begin{tabular}{ccccccc}
$\omega_\textrm{RF}/\omega_\textrm{rev}$ & n 
& $\Delta\theta/2\pi=0.1$
& $0.2$
& $0.3$
& $0.4$
& $0.5$
\\
\colrule
0.316 & 4 & \textcolor{red}{1.2} & \textcolor{red}{2.4} & \textcolor{red}{3.6} & \textcolor{red}{4.8} & \textcolor{red}{6}\\
-0.684 & 5 & \textcolor{red}{0.6} & \textcolor{red}{1.2} & \textcolor{red}{1.8} & \textcolor{red}{2.4} & \textcolor{black}{3}\\
1.316 & 3 & \textcolor{black}{1} & \textcolor{red}{2} & \textcolor{black}{3} & \textcolor{red}{4} & \textcolor{black}{5}\\
-1.684 & 6 & \textcolor{red}{0.8} & \textcolor{red}{1.6} & \textcolor{red}{2.4} & \textcolor{red}{3.2} & \textcolor{red}{4}\\
2.316 & 2 & \textcolor{red}{0.8} & \textcolor{red}{1.6} & \textcolor{red}{2.4} & \textcolor{red}{3.2} & \textcolor{red}{4}\\
-2.684 & 7 & \textcolor{black}{1} & \textcolor{red}{2} & \textcolor{black}{3} & \textcolor{red}{4} & \textcolor{black}{5}\\
3.316 & 1 & \textcolor{red}{0.6} & \textcolor{red}{1.2} & \textcolor{red}{1.8} & \textcolor{red}{2.4} & \textcolor{black}{3}\\
-3.684 & 8 & \textcolor{red}{1.2} & \textcolor{red}{2.4} & \textcolor{red}{3.6} & \textcolor{red}{4.8} & \textcolor{red}{6}\\
4.316 & 0 & \textcolor{red}{0.4} & \textcolor{red}{0.8} & \textcolor{red}{1.2} & \textcolor{red}{1.6} & \textcolor{red}{2}\\
-4.684 & 9 & \textcolor{red}{1.4} & \textcolor{red}{2.8} & \textcolor{red}{4.2} & \textcolor{red}{5.6} & \textcolor{black}{7}\\
5.316 & -1 & \textcolor{red}{0.2} & \textcolor{red}{0.4} & \textcolor{red}{0.6} & \textcolor{red}{08} & \textcolor{black}{1}\\
-5.684 & 10 & \textcolor{red}{1.6} & \textcolor{red}{3.2} & \textcolor{red}{4.8} & \textcolor{red}{6.4} & \textcolor{red}{8}\\
6.316 & -2 & \textcolor{red}{0} & \textcolor{red}{0} & \textcolor{red}{0} & \textcolor{red}{0} & \textcolor{red}{0}\\
-6.684 & 11 & \textcolor{red}{1.8} & \textcolor{red}{3.6} & \textcolor{red}{5.4} & \textcolor{red}{7.2} & \textcolor{black}{9}\\
\end{tabular}
\end{ruledtabular}
\end{table}

Table~\ref{tab:condition3a} shows the mitigated and unmitigated RF knockout conditions that satisfy Eqs.~(\ref{eq:mitigation21}) and (\ref{eq:mitigation22}) for various positions of the second cavity.

\section{\label{sec:measurement} Measurements at ISIS}

\subsection{\label{sec:measurement_exp-setup} Experimental setup}

While beam stacking is only possible in Fixed Field machines, the effects of the beam stacking RF programme can still be studied with the ISIS synchrotron. Key ISIS parameters relevant for the RF knockout experiments are given in table \ref{tab:isis_params}.

During normal operations the magnetic field in the ISIS accelerator varies according to the beam's momentum. However, the machine can be set to a `storage ring mode' where the magnets produce a constant magnetic field at the field strength for the injection energy of roughly \SI{70}{\mega\electronvolt}.

\begin{table}[h!]
\centering
\caption{Relevant ISIS parameters for RF knockout in storage ring mode.}
\label{tab:isis_params}
\begin{tabular}{l c}

\midrule
Injection energy & \SI{70}{\mega \electronvolt} \\
Injection revolution frequency   & \SI{0.675}{\mega \hertz} \\

Horizontal tune & 4.316 \\
Dispersion at $h=2$ cavities & \SI{3.3}{\meter} \\ 
Operational beam intensity & $\sim$ $2.5\times10^{13}$ \\
Typical intensity in \\RF knockout experiments & $\sim$ $2\times10^{12}$ \\
Incoherent space charge tune shift & $\sim$ -0.01 \\
\bottomrule
\end{tabular}
\end{table}

With a coasting beam stored at the ISIS injection momentum, a beam stacking RF programme can be imitated by sweeping the RF frequency as if it were accelerating a second beam during stacking.

\begin{figure}[htb]
    \centering

    \includegraphics[width=\columnwidth]{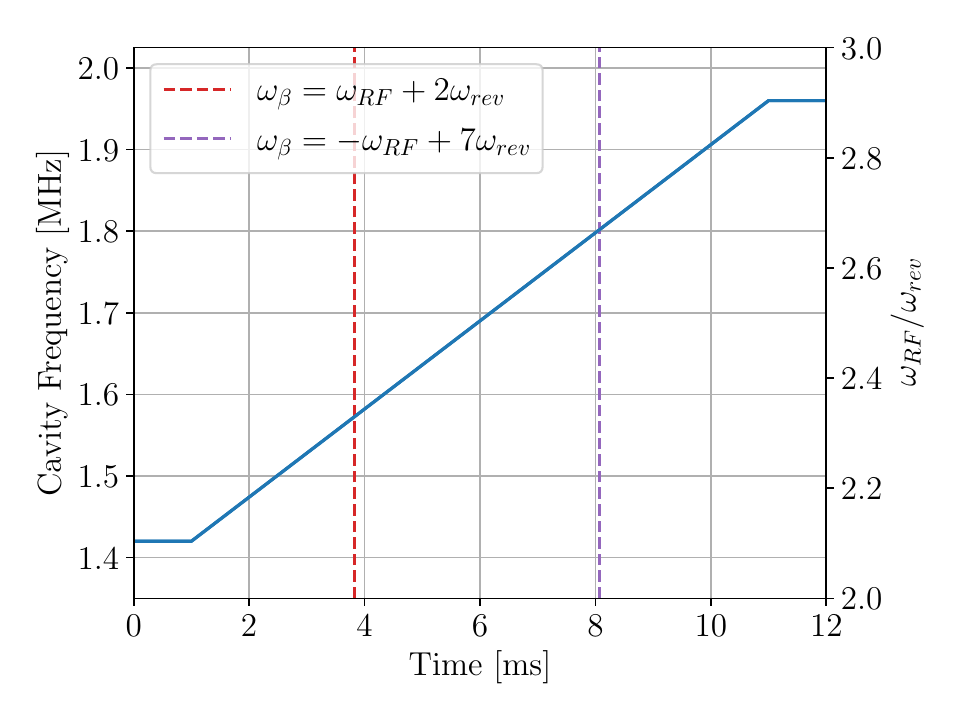}
    \caption{An example radio frequency cavity programme used to study RF knockout at ISIS. In this example the cavity frequency ramps from \SI{1.42}{\mega\hertz} to \SI{1.96}{\mega\hertz} or $2.1$ to $2.9$ times the coasting beam revolution frequency. The RF knockout conditions ($n=2$ and $n=7$) in this frequency range are indicated by the dashed lines.}
    \label{fig:RF_ramp_example}
\end{figure}

An example RF sweep is shown in Fig.~\ref{fig:RF_ramp_example}. The programme sweeps through nearly a whole RF harmonic to mimic the effect of the RF system accelerating a second, incoming beam on a stored coasting beam during the beam stacking process.

Operating the RF programme in Fig.~\ref{fig:RF_ramp_example} while a coasting beam is stored at the ISIS injection energy recreates the conditions during beam stacking in an FFA as shown in Fig.~\ref{fig:beam_stacking_example_program}. However, it is noted that the cavities will ramp through a frequency greater than the coasting beam revolution frequency. Whereas in the conventional beam stacking case the frequency of the accelerating second beam is less than that of the stored coasting beam. This difference will not affect the physics of the RF knockout, it only means that a different RF knockout condition will be encountered.

The RF programme ramps from $2.1$ to $2.9$ times the coasting beam revolution frequency. This range is within the normal operating parameters of the ISIS cavities and therefore ensures that the voltage across the cavity gap is as stable as possible over the ramp.

\subsection{Confirmation of RF knockout}

The current of the coasting beam was recorded while running the RF ramp shown in Fig.~\ref{fig:RF_ramp_example}. 

\begin{figure}[htb]
    \centering

    \includegraphics[width=\columnwidth]{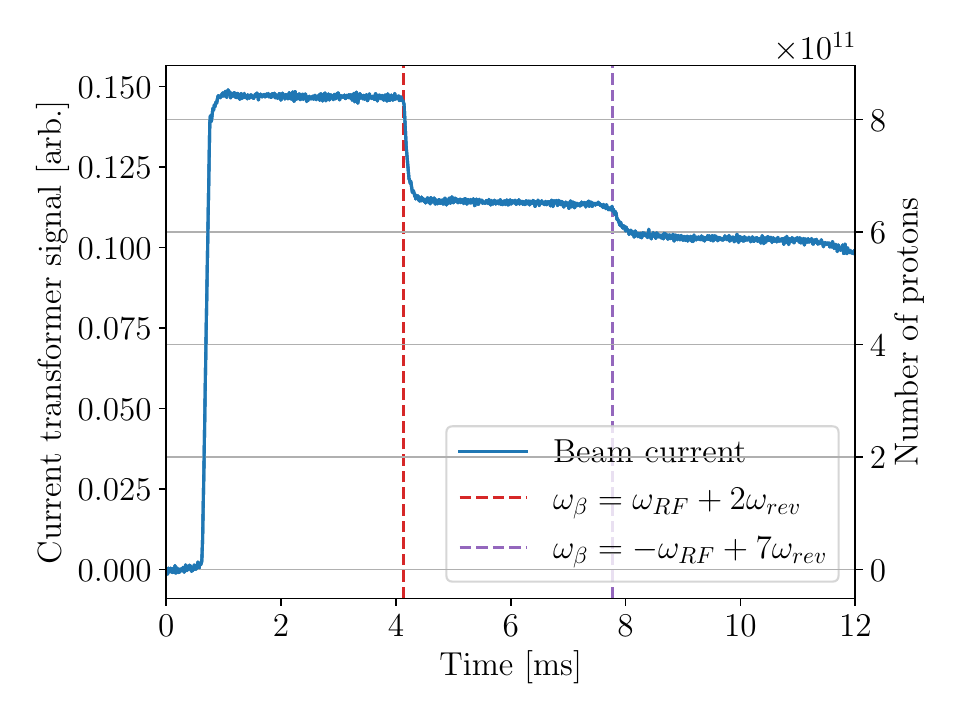}
    \caption{Beam current showing loss resulting from two RF knockout resonances.}
    \label{fig:RFKO_confirmation}
\end{figure}

In this particular frequency range, there are two frequencies where the condition for RF knockout is met and two corresponding drops in the beam current, as shown in Fig.~\ref{fig:RFKO_confirmation}.

The displacement given to the beam during RF knockout can be seen with a beam position monitor (BPM). The ISIS BPMs consist of two capacitive pickups above and below the beam for a vertical position monitor and to the inside and outside of the ring for a horizontal monitor. 

\begin{figure}[htb]
    \centering

    \includegraphics[width=\columnwidth]{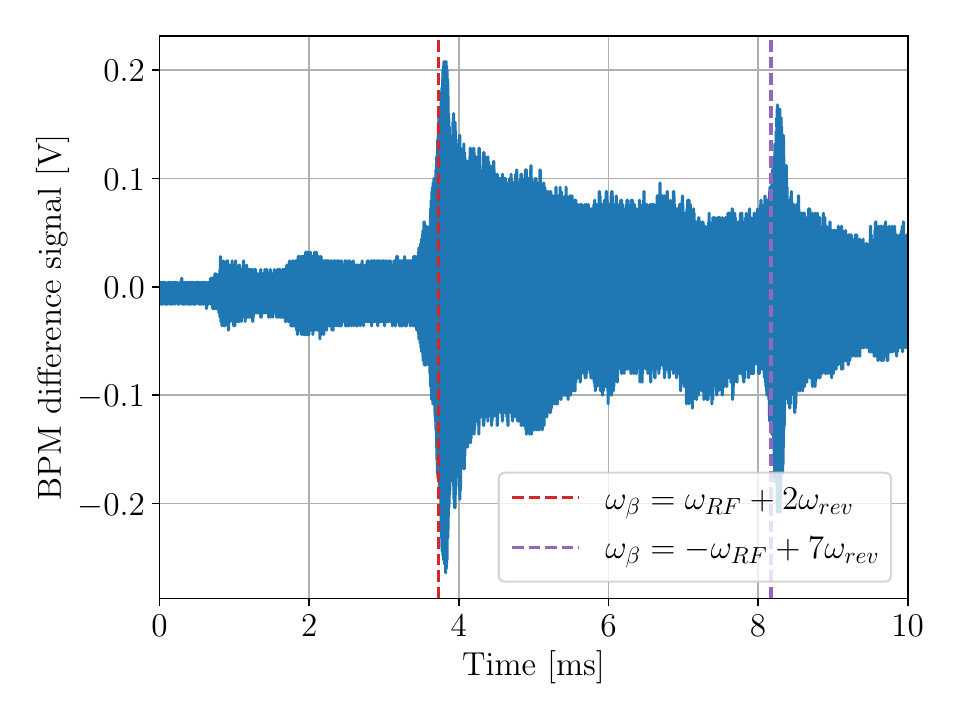}
    \caption{The difference signal from a horizontal beam position monitor during a 10 ms RF ramp between \SI{1.42}{\mega\hertz} and \SI{1.96}{\mega\hertz}. The beam's revolution frequency was \SI{0.675}{\mega\hertz}. The RF passes through two RF knockout conditions at \SI{4}{\milli\second} and \SI{8}{\milli\second} resulting in an increase in betatron oscillation amplitude.}
    \label{fig:RFKO_confirmation_BPM_diff}
\end{figure}

When the beam experiences a large displacement from RF knockout, the resulting horizontal betatron oscillation can be seen on the BPM difference signal. Figure \ref{fig:RFKO_confirmation_BPM_diff} shows the BPM difference signal over a 10 ms RF ramp, there are two increases in amplitude at approximately 4 and 8 ms indicating a large kick to the beam when the RF knockout condition is satisfied.

Since the beam remains un-bunched throughout the RF ramp, there is no discernible signal on the sum signal of the two BPM plates. Therefore, it is not possible to measure an absolute position variation of the beam centroid from RF knockout. However, the difference signal can still be used to infer a relative change in beam position and quantify an increase in beam amplitude.

\begin{figure}[htb]
    \centering
    \includegraphics[width=\columnwidth]{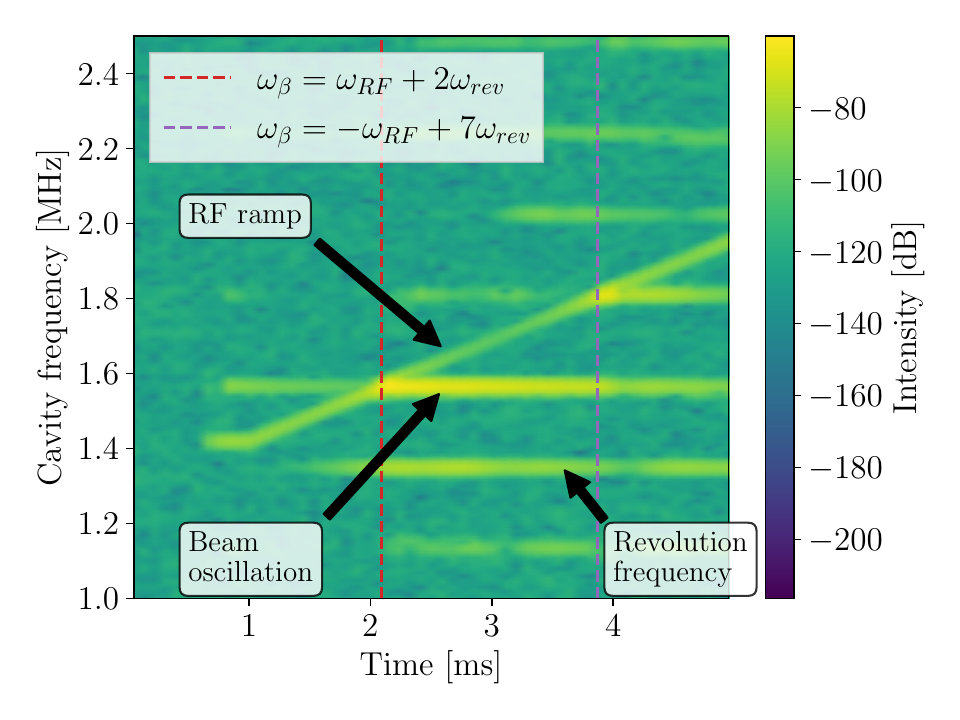}
    \caption{Spectrogram of the difference signal from a beam position monitor during RF knockout. The revolution frequency and betatron oscillation caused by the RF knockout displacements are labelled. The RF cavity induces noise on the beam position monitor which can be seen in the spectrogram.}
    \label{fig:RFKO_confirmation_spectrogram}
\end{figure}

Figure \ref{fig:RFKO_confirmation_spectrogram} shows a spectrogram of the BPM difference signal for a \SI{4}{\milli\second} RF ramp. When the RF knockout condition is met, a betatron oscillation begins at the RF knockout frequency, the betatron oscillation also appears as sidebands of the beam's revolution frequency.

\subsection{Parameter dependence}

The amplitude increase from passing through an RF knockout resonance is expected to vary linearly with the RF voltage and with the inverse square root of the crossing speed. To confirm that Eq.~(\ref{eqn:RFKO_amplitude}) holds true, the RF ramp time and RF voltage were scanned using the same experimental setup explained in Sec.~\ref{sec:measurement_exp-setup}.

For each ramp time and RF voltage, the BPM difference signal was acquired throughout the RF ramp. The amplitude of the betatron oscillation excited by passing through the RF knockout resonance was quantified by taking the Fourier transform of the BPM difference signal. A digital bandpass filter was then applied to isolate a specific harmonic of the betatron oscillation. The height of the peak was taken as a relative measurement of the change in beam amplitude from passing through the resonance.

\begin{figure}[htb]
    \centering

    \includegraphics[width=\columnwidth]{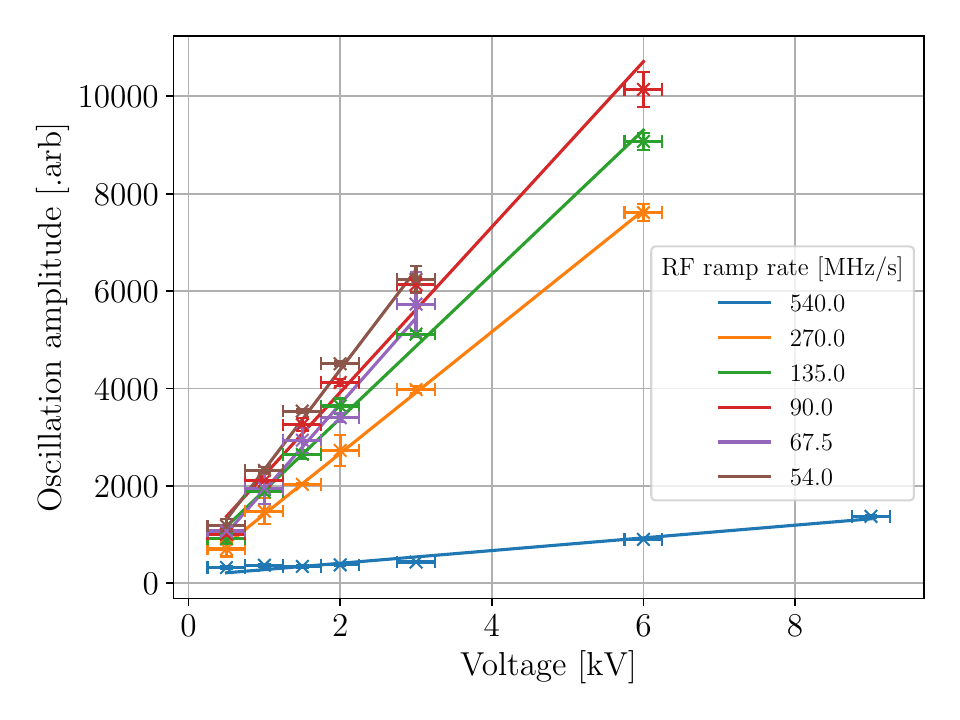}
    \caption{The relative amplitude change of the beam after passing through an RF knockout resonance for various RF ramp rates. The betatron oscillation amplitude is expected to vary linearly with the cavity voltage. }
    \label{fig:characterisation-FFT}
\end{figure}

The change in beam amplitude as measured from the Fourier transform is shown in Fig.~\ref{fig:characterisation-FFT}. The linear fit includes both the horizontal and vertical error bars.

The error in the oscillation amplitude is the standard error on the mean of three repeated measurements. The error on the RF gap voltage seen by the beam is estimated to be around \SI{250}{\volt}.

In the \SI{1}{\milli\second} ramp case the line is a poorer fit to the data as the fit is further from the estimated errors. Confidence in the model would be improved by taking additional repeats of this data but this was not possible with the limited machine time available.

\begin{figure}[htb]
    \centering
    \includegraphics[width=\columnwidth]{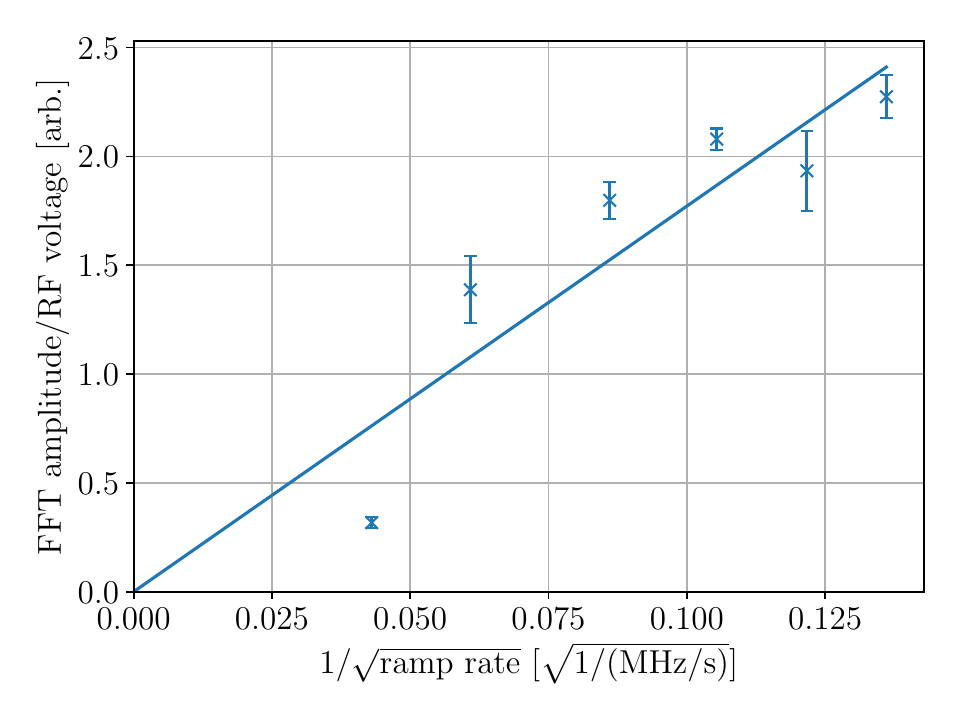}
    \caption{
    The amplitude of the betatron oscillation versus $1/\sqrt{\textrm{ramp rate}}$ of the RF ramp. The data has been scaled according to the RF voltage applied.}
    \label{fig:characterisation-FFT-ramp-rate}
\end{figure}

From Eq.~(\ref{eqn:RFKO_amplitude}), the change in beam amplitude is also expected to vary according to the inverse square root of the rate of frequency change. Figure~\ref{fig:characterisation-FFT-ramp-rate} shows the amplitude gain as a function of RF ramp time.

To assess if the experimental data is consistent with Eq.~(\ref{eqn:RFKO_amplitude}), the data is shown plotted as amplitude versus $1/\sqrt{\textrm{ramp rate}}$ in Fig.~\ref{fig:characterisation-FFT-ramp-rate}. The data should follow a linear relationship, with the gradient of the line representing the constant term in Eq.~(\ref{eqn:RFKO_amplitude}). Unfortunately, the numerical value of the constant term cannot be cross checked with the gradient of the fit as the beam position measurement is relative. 

Fitting a line to the logarithm of the betatron amplitude versus the logarithm of the ramp rate allows for the calculation of the exponent of the gradient, $\alpha$ in Eq.~(\ref{eqn:RFKO_amplitude}). The fit to the log data implies an exponent of $-0.78$ rather than the $-0.5$ expected from an inverse square root relationship. More repeats are needed to conclude that the amplitude varies with the inverse square root of the ramp rate.

Only data for which there was no significant beam loss is shown in Figs.~\ref{fig:characterisation-FFT}~and~\ref{fig:characterisation-FFT-ramp-rate}. If the beam intensity changes significantly as a result of the RF knockout then that will affect the amplitude measurement. To make sure only beams of similar intensity are compared, any datasets with clear beam loss on the current monitor are excluded. Significant loss appeared for ramp times of \SI{2}{\milli\second} or greater for voltages larger than \SI{6}{\kilo\volt}. For the \SI{6}{\milli\second} and \SI{10}{\milli\second} ramp times, only data up to \SI{3}{\kilo\volt} was included.

\subsection{\label{Mitigation_test}Mitigation test}

For beam stacking to be a viable technique in a high intensity FFA, beam loss due to RF knockout must be mitigated or avoided completely. In a ring with a single cavity, no mitigation or cancellation is possible. 

As introduced in Sec.~\ref{sec:RF knock-out} some of the resonance lines can be cancelled by placing cavities symmetrically throughout the ring. Alternatively, RF knockout can be mitigated by locally cancelling the displacements over a single turn. Here, both schemes are tested and compared experimentally.

\subsubsection{\label{sec:local_cancellation}Local cancellation with three cavities}

The diagram in Fig.~\ref{fig:3_cavity_local_cancellation} is exaggerated for clarity, in the ISIS case the two arc segments lie on the same radius and the angle between the vertical axis and the two dashed lines will be equal.

The RF cavity gap voltages required to cancel RF knockout can be calculated using the diagram. When $\nu_1 = \nu_2$, the amplitudes of the 3 cavities are related by;

\begin{equation}
    \begin{aligned}
        a_2 = 2 a_1 \text{cos}(\nu); \text{  where,}\\
        a_1 = a_3 \text{ and } \nu = \nu_1 = \nu_2.
    \end{aligned}
    \label{eqn:local_cancellation_eqn}
\end{equation}

\begin{figure}[htb]
    \centering
    \includegraphics[width=\columnwidth]{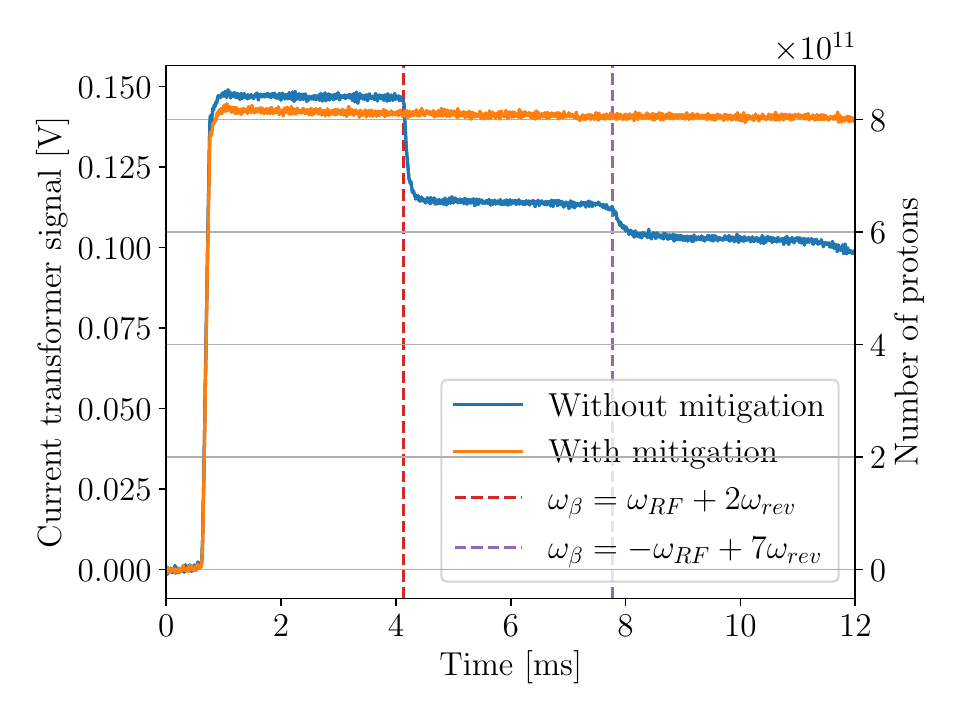}
    \caption{The beam current without and with the local cancellation mitigation method.}
    \label{fig:3_cav_mitigation_CT_comparison}
\end{figure}

For the phase advance between adjacent cavities in the ISIS ring, this leads to a voltage ratio across the 3 cavities of $0.55:1.0:0.55$. To confirm that this mitigation strategy works in practice this voltage ratio was applied to 3 of the ISIS cavities in consecutive superperiods. Otherwise, the experimental setup was similar to Sec.~\ref{sec:measurement_exp-setup} and the same frequency sweep shown in Fig.~\ref{fig:RF_ramp_example}, was used. The effect of the mitigation strategy on the coasting beam current is shown in Fig.~\ref{fig:3_cav_mitigation_CT_comparison}. The beam loss caused by RF knockout is completely eliminated to within the minimum sensitivity of the current transformer. 

\begin{figure}[htb]
    \centering
    \includegraphics[width=\columnwidth]{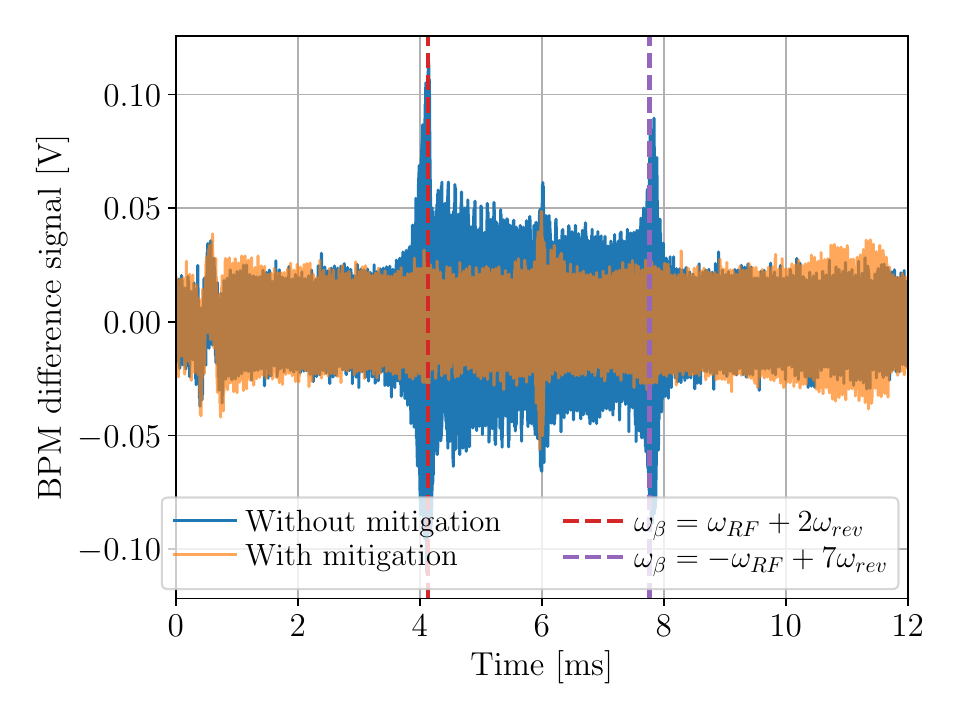}
    \caption{The beam position monitor difference signal without and with the local cancellation mitigation method. Both RF knockout conditions are suppressed by cancelling the displacements in a single turn with three successive cavities.}
    \label{fig:3_cav_mitigation_BPM_comparison}
\end{figure}

Although there was no observable beam loss with this mitigation strategy, the beam may still have received a kick from the RF knockout resonance; if present the kick should be visible in the BPM difference signal. Figure~\ref{fig:3_cav_mitigation_BPM_comparison} shows the BPM difference signals, with and without the mitigation strategy, overlaid on each other. In the unmitigated RF knockout the betatron oscillation can be seen at the two points where the RF knockout condition was met. In the data with mitigation, there is no identifiable oscillation indicating that the kick displacements were cancelled as predicted in Fig.~\ref{fig:3_cavity_local_cancellation}.

At around \SI{6}{\milli\second} in Fig.~\ref{fig:3_cav_mitigation_BPM_comparison}, a small peak appears in both the case with and without mitigation. The peak occurs exactly halfway through the RF ramp, when the RF frequency is $2.5$ times the revolution frequency. The peak is the result of some transient bunching of the beam from the $5^\text{th}$ harmonic of the RF frequency. The bunching quickly dissipates as the RF frequency moves past $2.5$ times the revolution frequency.

\subsubsection{Global cancellation with symmetric cavities}

In the previous section, the loss from RF knockout was eliminated by cancelling the displacements within a single turn. While the local cancellation method is effective, it is not a particularly practical option as the acceleration of the incoming beam is compromised to enable the mitigation of RF knockout.

The global cancellation method described in Sec.~\ref{sec:mitigation_with_symmetric_cavities} does not affect the acceleration of the incoming beam and is independent of the phase offset between the cavities. 

To test this mitigation method, the frequency ramp in Fig.~\ref{fig:RF_ramp_example} was applied using two cavities on exact opposite sides of the ISIS ring, delivering equal voltages. Figure~\ref{fig:ISIS_h=2_cavity_diagram} shows the position of the fundamental cavities in the ISIS ring. According to Eq.~\ref{eqn:opposite_cavities_time_indep_term} in the special case of two cavities on opposite sides of the ring, all odd numbered modes will be cancelled. For the ISIS accelerator with its harmonic two RF system, the $n=2$ mode will be uneffected but the $n=7$ mode will be cancelled. It is important to note that unlike the local cancellation method, the RF knockout displacements are not cancelled out in a single turn. The odd number oscillations triggered by RF knockout are cancelled because the symmetry of the cavity arrangement removes the RF knockout driving force.

\begin{figure}[htb]
    \centering
    \includegraphics[width=\columnwidth]{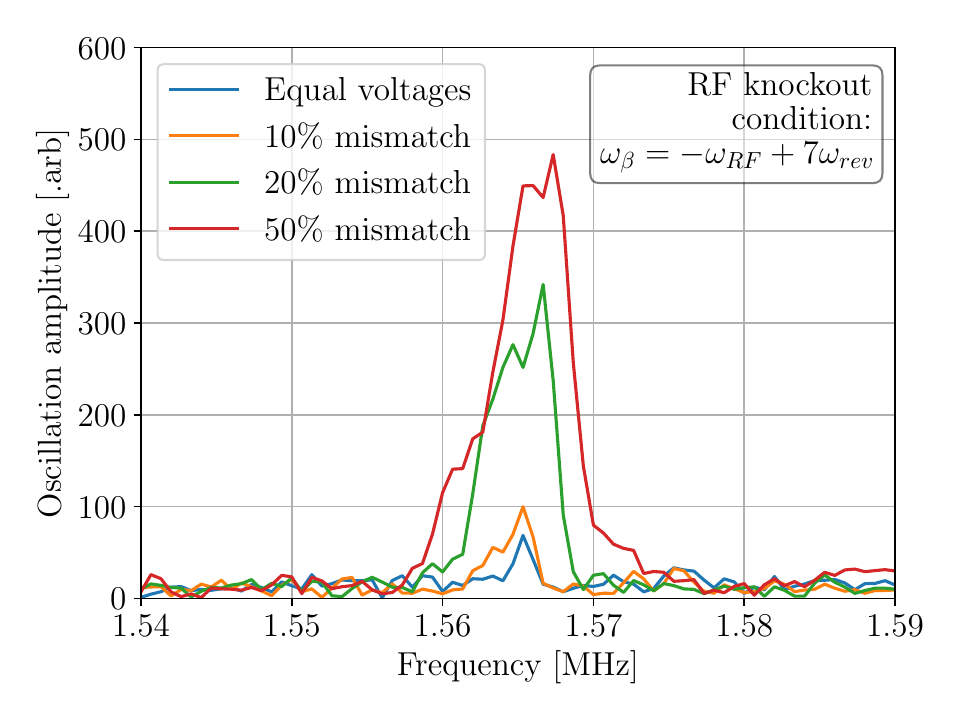}
    \caption{The betatron oscillation from the RF knockout for a mode number of seven is reduced with 2 cavities delivering equal voltages on opposite sides of the ISIS accelerator. As the voltage mismatch between the cavities is increased, the RF knockout displacements are no longer cancelled.}
    \label{fig:2_cav_mitigation_comparison}
\end{figure}

The amplitude of the $n=7$ oscillation for various cavity settings is shown in Fig.~\ref{fig:2_cav_mitigation_comparison}. With the two cavities set to equal voltages the oscillation induced by the RF knockout displacements is reduced, as shown by the blue line in Fig.~\ref{fig:2_cav_mitigation_comparison}. The voltage of one of the two cavities was then varied to deliver an unequal voltage to the beam. As the voltage mismatch increases, the symmetry is broken and the RF knockout kicks become more pronounced.

A detectable betatron oscillation persists even with the two cavities set to equal voltages. As the voltage mismatch increases the beam approaches the single cavity case and the resulting betatron oscillation grows. The comparatively small betatron oscillation with the cavities set to equal voltages is likely caused by the error on the voltage delivered by the cavities. Although there was an induced oscillation, there was no measurable loss on the ISIS beam current monitor in the equal voltage case.

\begin{table}[htb]
\caption{$\omega_\textrm{RF}/\omega_\textrm{rev}$ vs $\Delta\theta$ when $h=2$. When $\Delta\theta (h+n)/\pi$ or $\Delta\theta (-h+n)/\pi$ is an odd integer number, the knockout does not occur.}
\label{tab:condition3b}
\begin{ruledtabular}
\begin{tabular}{ccccccc}
$\omega_\textrm{RF}/\omega_\textrm{rev}$ & n 
& $\Delta\theta/2\pi=0.1$
& $0.2$
& $0.3$
& $0.4$
& $0.5$
\\
\colrule
2.316 & 2 & \textcolor{red}{0.8} & \textcolor{red}{1.6} & \textcolor{red}{2.4} & \textcolor{red}{3.2} & \textcolor{red}{4}\\
-2.684 & 7 & \textcolor{black}{1} & \textcolor{red}{2} & \textcolor{black}{3} & \textcolor{red}{4} & \textcolor{black}{5}\\
\end{tabular}
\end{ruledtabular}
\end{table}

Global cancellation with symmetric cavity positioning does not necessarily impose two cavities on the opposite sides of a ring.
That is only the special case of Eq.~(\ref{eqn:opposite_cavities_time_indep_term}) when $\Delta\theta=\pi$.
As shown in Fig.~\ref{fig:ISIS_h=2_cavity_diagram}, the ISIS accelerator has a 10 fold symmetry and the 6 RF fundamental cavities are installed at the identical position in each superperiod of \#2, \#3, \#4, \#7, \#8, \#9.
The combination of two cavities from different superperiods will yield the condition of $\Delta \theta/2\pi$ = 0.1, 0.2, 0.3, 0.4, and 0.5.
Table \ref{tab:condition3b} shows when the mitigation works for those combinations.

Figure \ref{fig:mitigation3_figures_appendix} in Appendix A shows that the RF knockout of the $n=7$ is suppressed when two cavities are apart by $\Delta\theta/2\pi=0.1, 0.3$ or $0.5$ as we expect from Table~\ref{tab:condition3b}.

\section{Discussion} \label{sec:discussion}

While the RF knockout mitigation schemes discussed in Sec.~\ref{Mitigation_test} can prevent beam loss during stacking, both methods have drawbacks. 

The local cancellation method cancels all RF knockout kicks but depends on the phase advance of the particle. It also compromises the accelerating RF programme, as both the phase and voltage of the cavities must change to enable the local cancellation. The requirement to change the phase and voltage to enable the local cancellation leads to an extra complexity. RF knockout occurs in the stored coasting beam at the top energy of the FFA; therefore any mitigation strategy must avoid RF knockout in the stored beam while maintaining the acceleration of the incoming beam. Here, the accelerating voltage has already been reduced to cancel the RF knockout kicks but furthermore, the cavities should be phased such that the beam receives the intended displacement from the cavity. This leads to a unique situation where the cavities need to be phased to perfectly cancel RF knockout while at the same time they need to be phased to maintain acceleration of the incoming beam. It is not possible to achieve both of these at the same time so an imperfect phase setting must be used for either the acceleration or the mitigation. 

To assess the sensitivity of the local cancellation to a phasing error, the experiment in Fig.~\ref{fig:3_cav_mitigation_CT_comparison} was repeated for a range of relative phases between the cavities. 
The cavity phase offset refers to the RF phase angle between successive cavities. Normally, when accelerating a beam that is synchronised with the RF waveform, the cavity phase offset is set equal to the RF phase angle between the cavities, $72\degree$. During RF knockout, the cavity frequency is not synchronised with the stored beam therefore an adjusted phase offset should be used to ensure full cancellation of the displacements.

\begin{figure}[htb]
    \centering
    \includegraphics[width=\columnwidth]{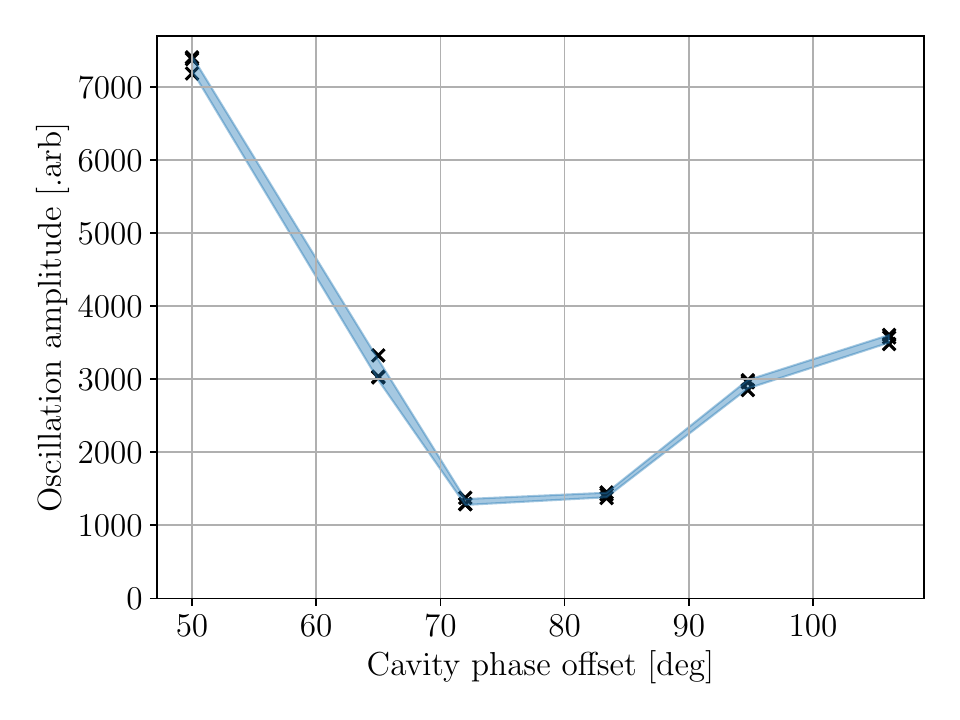}
    \caption{The amplitude of the betatron oscillation triggered by RF knockout varies with a phasing error between cavities. Three repeats for each phase are shown. The shaded area indicates the standard error on the mean of the three repeats.}
    \label{fig:3_cav_mitigation_phase_scan}
\end{figure}

In this experiment, RF knockout occurs at cavity frequencies $2.319$ ($n=2$ mode) and $2.681$ ($n=7$ mode) times the revolution frequency. The cavity phase offset must be adjusted to account for the cavity frequency at which the RF knockout condition is met. The cavity phase offset is given by; $\frac{\omega_\textrm{RF}}{\omega_\textrm{rev}} \frac{2 \pi}{N}$ where $N$ is the number of of superperiods (for ISIS, $N=10$).

Taking the $n=2$ RF knockout, a phase offset of roughly $83.7\degree$ matches the phase of RF to the beam at a cavity frequency of $2.319$ times the revolution frequency. Then, the combination of the 3 cavities should cancel the RF knockout displacements perfectly. Figure \ref{fig:3_cav_mitigation_phase_scan} shows the effect of changing the cavity phase offset. As the phase moves away from the ideal value, the displacements are no longer cancelled and an oscillation of the beam centroid can be detected.

Using Eq.~(\ref{eqn:local_cancellation_eqn}) to calculate the cavity phase offset $\nu$ for the ISIS case---an $83\degree$ phase offset should perfectly cancel the RF knockout kicks and therefore prevent the amplitude growth from RF knockout. However, there is still a detectable oscillation at the horizontal tune frequency for the $83\degree$ offset.

Clearly the RF knockout displacements were not completely cancelled. The most likely cause of the persisting oscillation is the uncertainty in the RF voltages applied to the beam. The voltage produced at the cavity gap is accurate to around \SI{250}{\volt}. Ideally, the RF voltage across the three cavities should follow the ratio $0.55:1.0:0.55$ but in reality, it is probable that this ratio wasn't quite achieved, leading to a detectable oscillation even when the cavities were phased perfectly. 

The cable delays in the RF system were recently measured and adjusted; it is thought that the error in the RF waveform phase is small compared with the error on the voltage. 

Unlike the local cancellation method, the global cancellation with symmetrical cavities is independent of the phase advance and does not compromise the accelerating RF programme. Nevertheless, the global cancellation method has some disadvantages. It does not cancel every occurrence of RF knockout and requires specific cavity positions in the ring.

Beam stacking is particularly useful in a high intensity FFA \cite{BeamStackingPaper_PhysRevAccelBeams, FFA_CDR_ePub}; collective effects associated with a highly intense beam could be present before, during and after the stacking process. The RF knockout cancellation responds differently to a space-charge tune spread depending on the mitigation strategy.

In the local cancellation case, the elimination of RF knockout depends on the total displacements summing to zero in a single turn of the ring. The local cancellation method therefore depends on the phase advance of individual particles in the beam. In the high intensity scenario, the beam will have a space-charge tune spread and the diagram in Fig.~\ref{fig:3_cavity_local_cancellation} will no longer apply exactly to every particle as each particle's phase advance will be different and the effect of the local cancellation may be degraded.

The global cancellation by symmetric cavities mitigates some of the RF knockout kicks with a different principle. This method eliminates RF knockout by cancelling the displacements from the cavities. Since this method does not rely on the phase advance of an individual particle it should be unaffected by a space-charge tune spread. We therefore consider that the global cancellation method is the best option for a high intensity FFA. The RF knockout free frequency range can be increased by placing more cavities symmetrically around the ring. 

In the experimental studies of the local cancellation method a low intensity coasting beam of around $2\times10^{12}$ particles was used. The space-charge tune shift is almost non-existent at this intensity as the beam is un-bunched. During user cycles ISIS operates at an intensity around 10 times greater, but it is difficult to store this intensity in storage ring mode at the injection energy. High intensity experimental studies is one avenue for future research.

\section{\label{sec:conclusion} Conclusion}
RF knockout has been confirmed experimentally as the source of beam loss during beam stacking at the KURNS FFA. There are no dispersion free regions in a scaling FFA; therefore, for beam stacking to be a viable technique, the beam loss from RF knockout must be mitigated. We have shown experimentally that RF knockout can be mitigated with two techniques, local and global cancellation. The local cancellation method can cancel all RF knockout displacements but could also compromise the acceleration of beams during stacking. While the global cancellation does not cancel all RF knockout oscillation modes, the global cancellation method can create a RF knockout free frequency range large enough for a beam stacking RF programme to be executed without compromising beam acceleration. Furthermore, the global cancellation is independent of the particle tune making it a viable option for beam stacking at a high intensity FFA for a future Spallation Neutron Source.

\bibliographystyle{apsrev4-2}
\bibliography{apssamp}

\appendix
\section{RF knockout mitigation with non-symmetric RF cavities}
%\begin{figure*}[htb]
%    \centering
%    \includegraphics[width=\textwidth]{Figures/Appendix/non_symmetric_cavities_mitigation_time_domain.pdf}
%    \caption{The time domain data from a beam position monitor difference signal showing how the oscillation triggered by RF knockout is mitigated/unmitigated depending on the arrangement of RF cavities.}
%    \label{fig:mitigation3_figures_appendix}
%\end{figure*}

Experimental results on the RF knockout mitigation by two RF cavities are shown here.
Because of a 10-fold symmetry of the ISIS lattice, it is possible to test various values of $\Delta\theta$ in Eqs.~(\ref{eq:mitigation21}) and (\ref{eq:mitigation22}).

\begin{figure*}[p]                  
    \centering
    \begin{subfigure}[t]{0.48\textwidth}
        \centering
        \includegraphics[width=\textwidth]{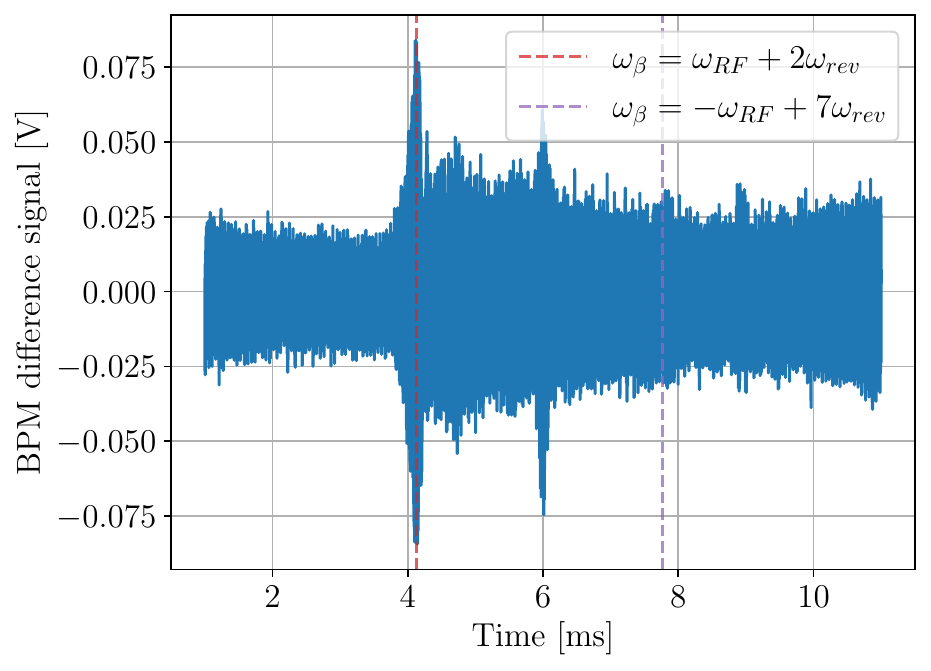}
        \caption{$\Delta\theta/2\pi=0.1$ (Cavities 2 and 3). mitigated}
    \end{subfigure}
    \hfill
    \begin{subfigure}[t]{0.48\textwidth}
        \centering
        \includegraphics[width=\textwidth]{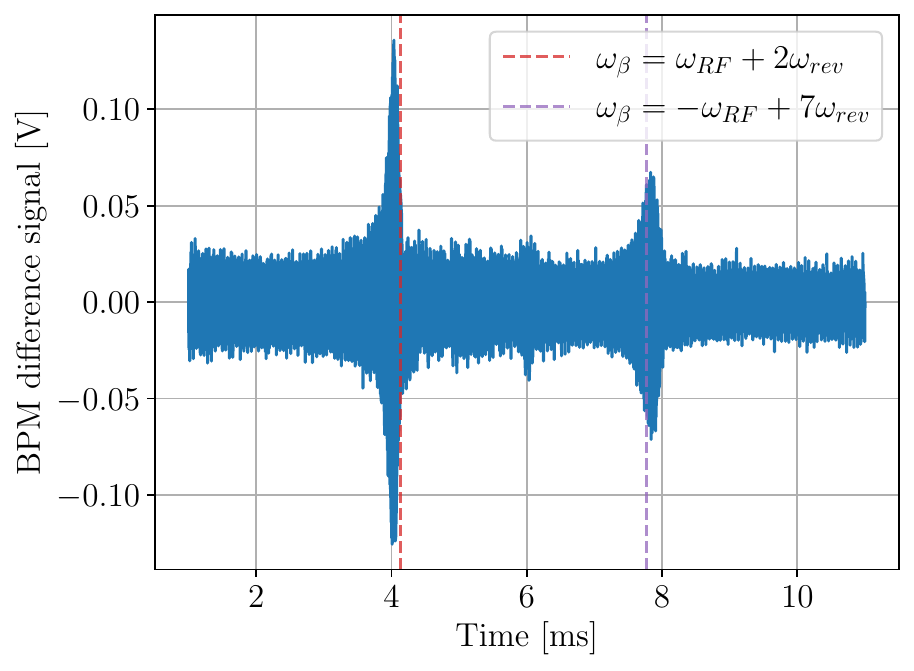}
        \caption{$\Delta\theta/2\pi=0.2$ (Cavities 2 and 4). unmitigated}
    \end{subfigure}
    \vspace{1em}
    \begin{subfigure}[t]{0.48\textwidth}
        \centering
        \includegraphics[width=\textwidth]{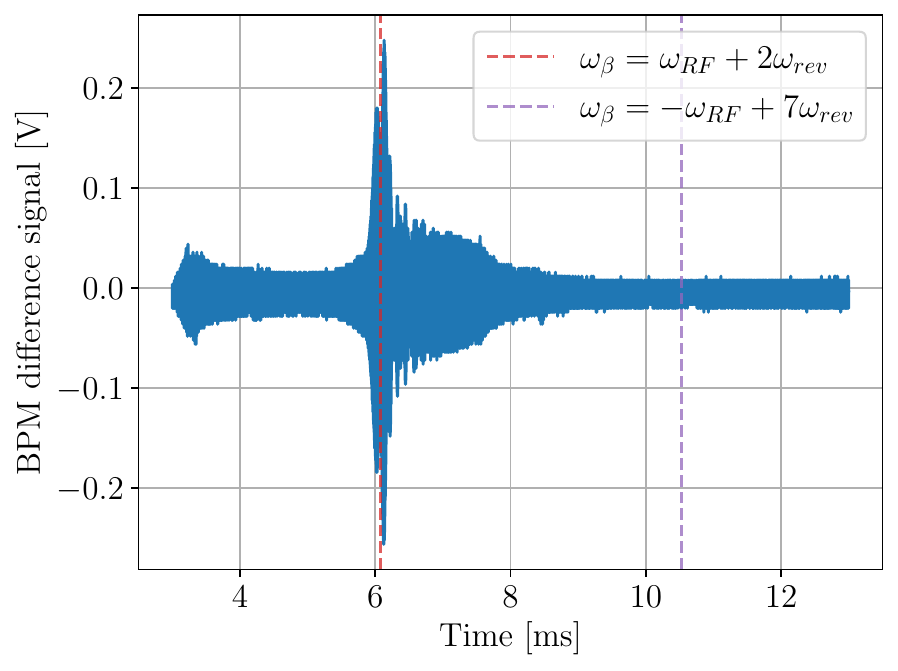}
        \caption{$\Delta\theta/2\pi=0.3$ (Cavities 7 and 4). mitigated}
    \end{subfigure}
    \hfill
    \begin{subfigure}[t]{0.48\textwidth}
        \centering
        \includegraphics[width=\textwidth]{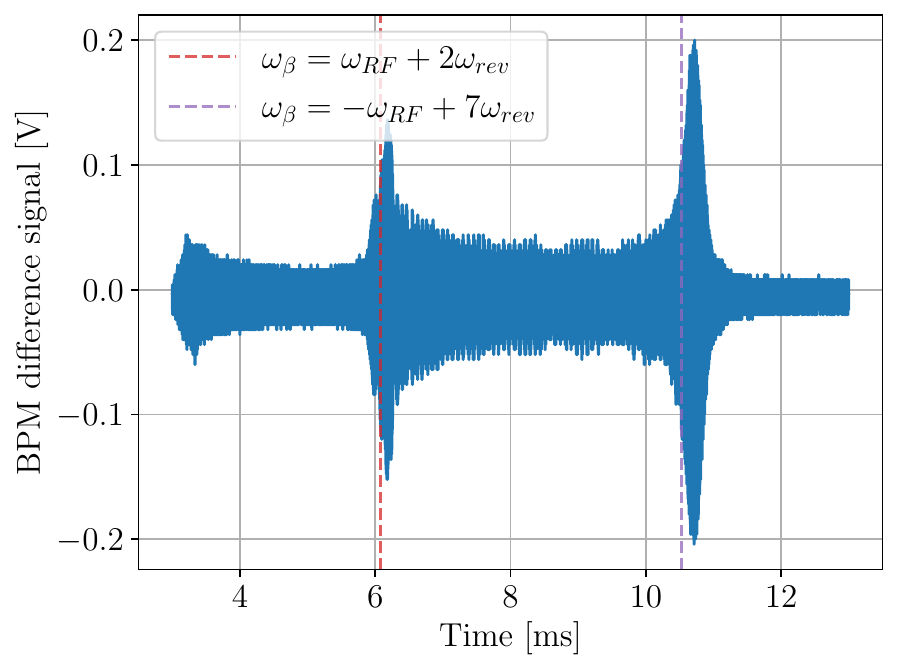}
        \caption{$\Delta\theta/2\pi=0.4$ (Cavities 7 and 3). unmitigated}
    \end{subfigure}
    \vspace{1em}
    \begin{subfigure}[t]{0.48\textwidth}
        \centering
        \includegraphics[width=\textwidth]{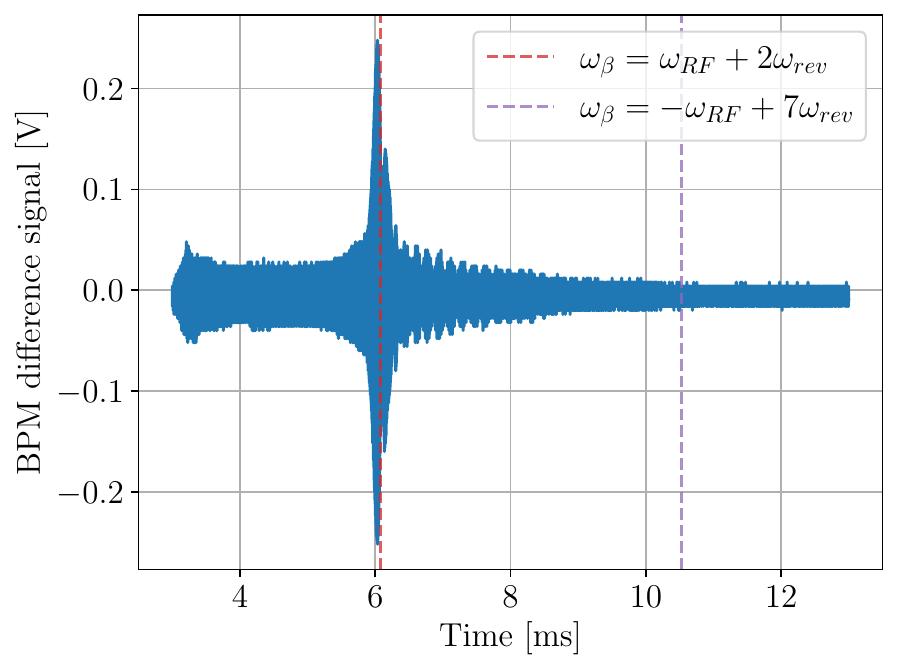}
        \caption{$\Delta\theta/2\pi=0.5$ (Cavities 7 and 2). mitigated}
    \end{subfigure}
    \hfill{}

    \caption{The time domain data from beam position monitor difference signals showing how the oscillation triggered by RF knockout is mitigated/unmitigated depending on the arrangement of RF cavities.}
    \label{fig:mitigation3_figures_appendix}
\end{figure*}

%\begin{figure}[h]
%	\centering
%	\subfloat[][$\Delta\theta/2\pi=0.1$ with RF \#2 and \#3.]{
%	\includegraphics[width=0.48\linewidth]{Figures/Measurement/RF2and3_6.pdf}\hfill
%	\label{fig:sim:phasesp_a}
%	}
%	\subfloat[][$\Delta\theta/2\pi=0.2$ with RF \#2 and \#4.]{
%	\includegraphics[width=0.48\linewidth]{Figures/Measurement/RF2and4_7.pdf}
%	\label{fig:sim:phasesp_b}
%	}\\
%	\subfloat[][$\Delta\theta/2\pi=0.3$ with RF \#4 and \#7.]{
%	\includegraphics[width=0.48\linewidth]{Figures/Measurement/RF4and7_42.pdf}\hfill
%	\label{fig:sim:phasesp_c}
%    }
%	\subfloat[][$\Delta\theta/2\pi=0.4$ with RF \#3 and \#7.]{
%	\includegraphics[width=0.48\linewidth]{Figures/Measurement/RF3and7_38.pdf}
%	\label{fig:sim:phasesp_d}
%    }\\
%	\subfloat[][$\Delta\theta/2\pi=0.5$ with RF \#2 and \#7.]{
%	\includegraphics[width=0.48\linewidth]{Figures/Measurement/RF2and7_34.pdf}
%	\label{fig:sim:phasesp_d}
%    }
%\caption{Horizontal beam position monitor difference signal.}
%\label{fig:mitigation3_figures}
%end{figure}

\begin{acknowledgments}

We wish to acknowledge the operations team of the ISIS accelerator, especially H.~Cavanagh and the ISIS Crew. Thanks also to R.~Williamson for his comments on the~manuscript.

\end{acknowledgments}

\end{document}